\documentclass[journal]{IEEEtran}

\usepackage{hyperref}
\usepackage{graphicx}
\usepackage{subcaption}
\usepackage{comment}
\usepackage[backend=biber, sorting=none]{biblatex}
 \usepackage[acronym, nonumberlist,nohypertypes={acronym}]{glossaries}
\newacronym{llm}{LLM}{Large Language Model}
\newacronym{ocr}{OCR}{Optical Content Recognition}
\newacronym{rpa}{RPA}{Robotic Process Automation}
\newacronym{ai}{AI}{Artificial Intelligence}
\newacronym{crm}{CRM}{Customer Relationship Management}
\newacronym{erp}{ERP}{Enterprise Resource Planning}
\newacronym{hris}{HRIS}{Human Resource Information System}
\newacronym{mcp}{MCP}{Model Context Protocol} % load the acronyms file

\usepackage{cleveref}

\begin{document}

\title{Email as the Interface to Generative AI Models: Seamless Administrative Automation}

\author{
\IEEEauthorblockN{
Andr\'{e}s~Navarro\IEEEauthorrefmark{1}, 
Carlos~de~Quinto\IEEEauthorrefmark{1}, 
Jos\'{e}~Alberto~Hern\'{a}ndez\IEEEauthorrefmark{1} 
%D. Scano\IEEEauthorrefmark{2}, F. Cugini\IEEEauthorrefmark{3}, G. Mart\'{i}nez\IEEEauthorrefmark{1},
%\'{O}. Gonz\'{a}lez de Dios\IEEEauthorrefmark{4} %DD\IEEEauthorrefmark{3} and %EE\IEEEauthorrefmark{4}
}
\\
\IEEEauthorblockA{\IEEEauthorrefmark{1} Dept. Telematics Engineering, Universidad Carlos III de Madrid, Spain}\\
}

\maketitle

\begin{abstract}
This paper presents a novel architectural framework that integrates Large Language Models (LLMs) with email interfaces to automate administrative tasks, addressing critical accessibility barriers in enterprise environments. The proposed system integrates email communication channels with Optical Character Recognition (OCR) and intelligent automation, allowing non-technical administrative staff to delegate complex form-filling and document processing tasks through familiar interfaces. Our approach uses email body content as natural language prompts and attachments as contextual information, creating a seamless workflow that bridges the gap between sophisticated AI capabilities and practical usability. Empirical evaluation demonstrates that the system can complete complex administrative forms in under 8 seconds of automated processing, with human supervision reducing total staff time by a factor of three to four compared to manual workflows. The top-performing LLM correctly filled 16 out of 29 form fields, and 64\% relatively reduced the total cost per processed form to manual completion. These results establish email-based LLM integration as a viable and cost-effective approach for democratizing advanced automation in organizational settings, enabling widespread adoption without requiring specialized technical knowledge or extensive workflow modifications.
\end{abstract}

\begin{IEEEkeywords}
Generative AI; Large Language Models; Administrative Automation; Email Interfaces; Optical Character Recognition; Cost Reduction; Workflow Efficiency.
\end{IEEEkeywords}

\IEEEpeerreviewmaketitle

\section{Introduction}

The proliferation of \glspl{llm} has fundamentally transformed approaches to automating language-intensive tasks across diverse organizational contexts \autocite{10527275, 10.1145/3636534.3649379}. These sophisticated models demonstrate remarkable capabilities in processing natural language instructions and generating contextually appropriate responses, positioning them as powerful tools for reducing administrative burden and enhancing operational efficiency \autocite{Peddinti2023}. However, despite their impressive technical capabilities, the practical implementation of \glspl{llm} in everyday administrative workflows remains constrained by significant accessibility barriers, particularly for non-technical users who constitute the majority of administrative personnel in most organizations \autocite{Pesch_2025}.

Administrative tasks represent a substantial component of organizational operations, with employees frequently dedicating considerable time to repetitive processes including form completion, data entry, document processing, and information extraction \autocite{21068f9715904033900a91cad6b9f758}. Many of these tasks exhibit structured characteristics and heavy reliance on textual information, making them ideal candidates for automation through \gls{llm}-based approaches. Recent research has demonstrated the effectiveness of \glspl{llm} in clinical documentation, prior authorization processes, and various administrative functions within healthcare settings \autocite{10527275, ess2025ai}, suggesting broader applicability across organizational domains.

The challenge of accessibility presents the most significant barrier to widespread \gls{llm} adoption in administrative contexts \autocite{21068f9715904033900a91cad6b9f758}. While these models offer impressive capabilities for task automation, integrating them into the daily workflows of administrative staff without extensive technical expertise requires intuitive interfaces that align seamlessly with existing work practices. Traditional approaches to \gls{llm} deployment often necessitate specialized software, complex configuration procedures, or technical training that may be prohibitive for many administrative users \autocite{chen2025empiricalstudychallengesllm}.

Email systems represent the most ubiquitous communication platform among administrative employees across virtually all organizational settings \autocite{kong2023organizationalbulkemailsystems, josephs2024communicationnetworkdynamicslarge}. The familiarity and universal adoption of email interfaces suggest significant potential for bridging the accessibility gap between sophisticated \gls{llm} capabilities and practical administrative workflow integration. Research in email automation has demonstrated promising results in information extraction, classification, and response generation, indicating the viability of email-based approaches for more complex administrative task automation \autocite{Khare_2022, thiergart2021understandingemailsdraftingresponses}.

This paper proposes a comprehensive architectural framework that leverages email interfaces as the primary interaction mechanism for \gls{llm}-based administrative task automation. Our approach utilizes email body content as natural language prompts and attachments as contextual information, leveraging \gls{ocr} and data retrieval techniques, to enable administrative staff to delegate complex form-filling and document processing tasks through familiar communication channels. The framework integrates multiple technological components including email clients, \gls{ocr} modules, \glspl{llm}, and Python-based automation scripts to create a seamless workflow for administrative task completion.

\begin{figure}
    \centering
    \includegraphics[width=1\linewidth]{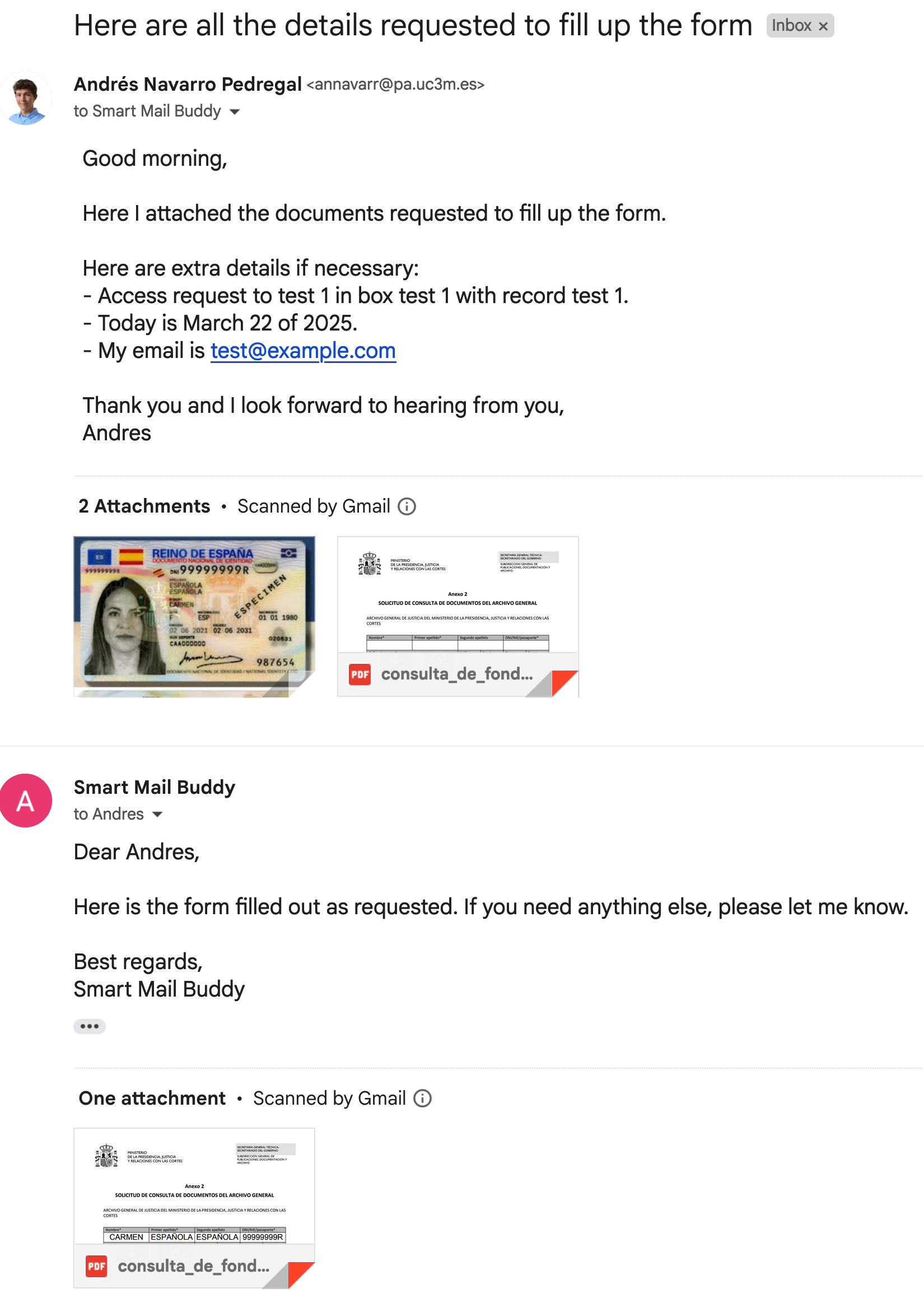}
\caption{Example email workflow for administrative automation: A user submits required details and scanned documents via email, and the automated system replies with the completed form attached, demonstrating seamless end-to-end processing using natural language instructions and document attachments.}
    \label{fig:email_example}
\end{figure}

The research contributions include: (1) an architectural framework combining data extraction of images with \gls{llm}-based automation for administrative tasks; (2) comprehensive evaluation across multiple administrative scenarios demonstrating significant efficiency improvements; (3) analysis of practical implementation considerations including cost-effectiveness, accuracy, and accessibility factors; and (4) identification of future research directions for expanding \gls{llm}-based administrative automation capabilities. See Figure \ref{fig:email_example} for reference.

\section{Related Work}

\subsection{Email-Based Automation Systems}

Google's Smart Reply system established foundational approaches for email automation through automated methods such as neural networks \autocite{kannan2016smartreplyautomatedresponse}. The system generates semantically diverse response suggestions that can serve as complete email responses, demonstrating the viability of email-based natural language processing for practical applications. However, Smart Reply focuses primarily on response generation rather than comprehensive task automation, indicating opportunities for extending email-based approaches to more complex administrative workflows.

Recent research in email automation has explored \gls{rpa} combined with \gls{ai} for comprehensive email handling and management \autocite{Khare_2022}. These systems demonstrate capabilities for automatic mailbox login through secured channels, email classification, attachment processing, and automated response generation. Our proposed framework builds upon these foundational capabilities while extending functionality to include complex form completion and document processing tasks.

\subsection{Form Filling and Document Processing Automation}

The evolution of form-filling automation has progressed from simple auto-fill capabilities to sophisticated context-aware systems \autocite{bose2019fieldlabelpredictionautofill, aveni2023omnifilldomainagnosticformfilling}. Traditional browser-based auto-fill systems predict field labels and automatically populate values based on previously filled forms, providing basic automation for repetitive data entry tasks. However, these systems typically operate within limited domains and lack the contextual understanding necessary for complex administrative scenarios.

OmniFill represents a significant advancement in domain-agnostic form filling through \gls{llm} integration \autocite{aveni2023omnifilldomainagnosticformfilling}. The system analyzes compound contextual demands and leverages LLM capabilities to infer suggestions for diverse form-filling tasks that were previously beyond the reach of traditional approaches. Our framework extends these capabilities by integrating email-based instruction processing with comprehensive document context analysis.

\subsection{Optical Character Recognition and Document Processing}

Modern \gls{ocr} technology has evolved from simple character recognition to sophisticated document processing systems capable of handling diverse formats and languages. Contemporary \gls{ocr} solutions utilize artificial intelligence and machine learning algorithms to continuously improve accuracy and adapt to different fonts, languages, and writing styles. Advanced systems, such as AWS Textract \autocite{amazon_textract} and Amazon Rekognition \autocite{amazon_rekognition}, provide comprehensive document analysis capabilities, extracting not only text but also understanding document structure, tables, and form fields. These cloud-based solutions offer high accuracy through deep learning technology and support for broad language recognition, making them suitable for enterprise-level document processing workflows. However, open-source alternatives like EasyOCR \autocite{easyocr} and TesseractOCR \autocite{tesseract2025} provide accessible options for developers, though they may not match the comprehensive feature sets of commercial solutions.

\section{Methodology}

\begin{figure}
    \centering
    \includegraphics[width=0.8\linewidth]{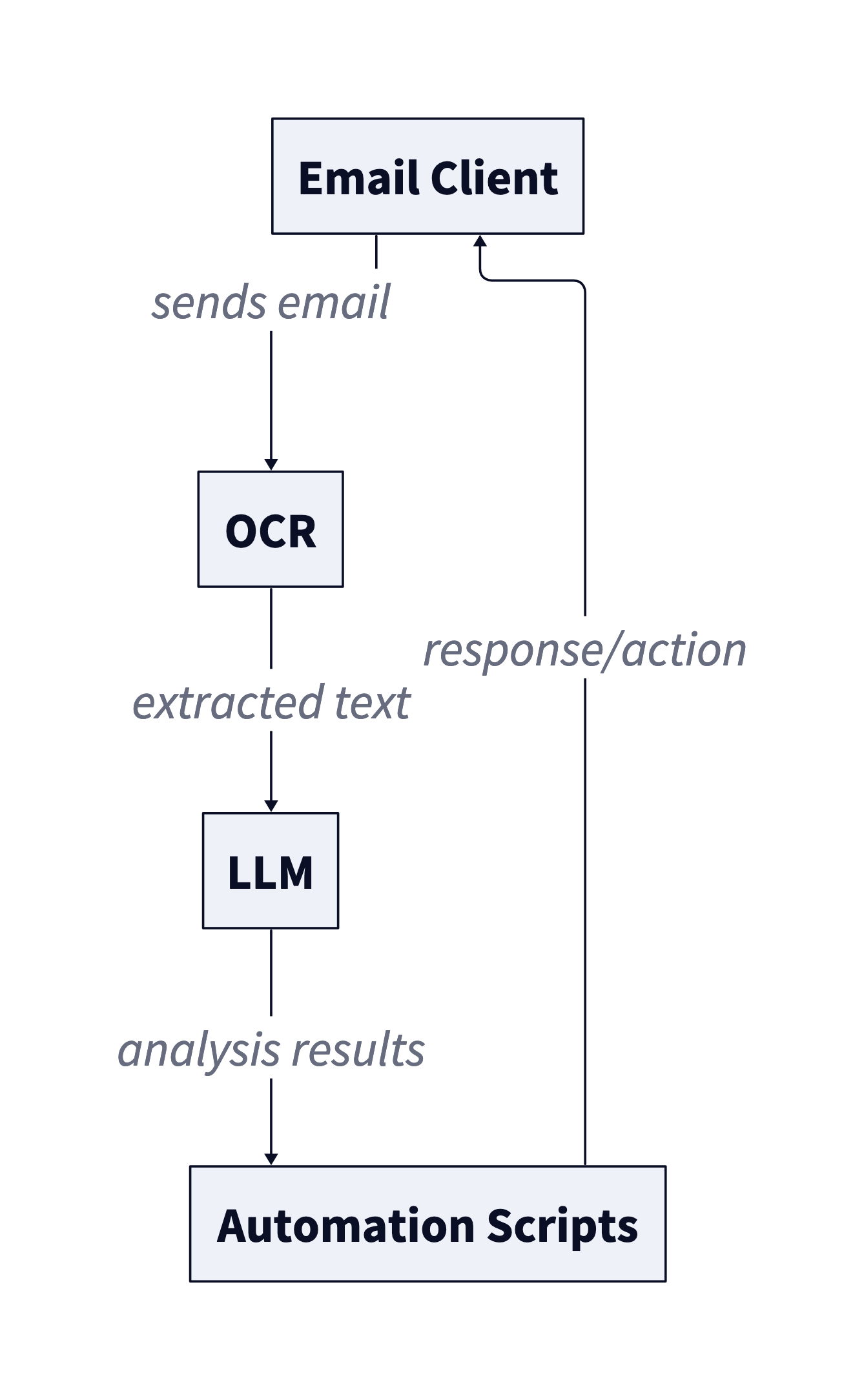}
\caption{System workflow architecture showing the integration of Email Client, OCR module, LLM, and automation scripts for administrative task automation.}
    \label{fig:workflow}
\end{figure}

Our proposed system integrates several key components to automate administrative tasks through email interfaces. The architecture consists of four primary components: an Email Client, \gls{ocr} module, a \gls{llm}, and Python-based automation scripts. \Cref{fig:workflow} presents a block diagram of the system architecture.

\subsection{Email Client}

The email client serves as the primary interface between the users and the automation system. It monitors a designated email inbox for incoming requests, processes new messages, extracts email body content as instructions, and handles attachments such as forms and supporting documents. The email client is implemented using the Amazon Workmail \autocite{amazonworkmail}.

This approach builds upon established email automation research, such as Google's Smart Reply system \autocite{kannan2016smartreplyautomatedresponse}, which generates semantically diverse suggestions that can be used as complete email responses. However, our system extends beyond response suggestions to enable the automation of complex tasks through the email channel.

\subsection{Optical Character Recognition (OCR)}

The \gls{ocr} component processes attached documents, converting image-based content into machine-readable text. This is crucial for handling scanned forms, certificates, and other documents that contain important information needed for form completion.

The \gls{ocr} module significantly reduces input errors. It accelerates the form processing workflow by automatically extracting relevant information from attached documents, which can then be effectively utilized by the \gls{llm} for administrative task completion.

\subsection{Large Language Model (LLM)}

The \gls{llm} is responsible for several critical functions, including understanding user intent from email body text, extracting relevant information from \gls{ocr}-processed documents, determining the appropriate administrative task workflow, generating accurate text for form completion, and providing explanations and clarifications in response emails. These capabilities work together to create a comprehensive intelligent system that can handle complex administrative workflows with minimal human intervention.

To improve performance on administrative tasks, we implement prompt engineering techniques and context optimization to help \gls{llm} focus on relevant information.

\subsection{Python Automation Framework}

The Python framework orchestrates the entire system, managing workflow between components and handling specialized tasks such as form field identification and completion, PDF manipulation for form filling, database interactions for verification or record-keeping, error handling and exception management, as well as performance monitoring and logging. These capabilities ensure that the system can handle the technical complexities of automated form processing while maintaining reliability and traceability throughout the workflow.

This component utilizes libraries such as PyPDF, pandas, and custom automation scripts to execute the technical aspects of form completion once the \gls{llm} has determined the appropriate content. The framework follows a structure similar to the Agent-S architecture \autocite{kulkarni2025agentsllmagenticworkflow}, which uses task-specific LLMs augmented with memory and environments for automating Standard Operating Procedures.

\section{Proof of concept}

To demonstrate the effectiveness of our proposed architecture, we implemented a proof-of-concept scenario representing common administrative tasks. This implementation showcases the system's versatility in handling various form types and requirements across different organizational contexts.

The proof of concept demonstrates autonomous form completion capabilities, where the system receives PDF forms as documents with instructions and reference information. The system analyzes form structure, identifies required fields, extracts relevant information from the provided context, and automatically populates forms with appropriate content. This approach ensures that the automated processing maintains the precision expected from manual completion.

The complete system workflow operates through a systematic approach ensuring seamless processing from request to delivery. When requests arrive, the system extracts the text as instruction prompts and processes documents through the \gls{ocr} module to extract textual content. The \gls{llm} receives instruction prompts and \gls{ocr}-extracted text as context, interprets requests and generates detailed completion plans. The Python automation framework executes these plans, filling form fields with extracted information.

This workflow enables administrative staff to delegate form-filling tasks by integrating with email clients and sending emails with the required information and attached forms, receiving completed documents without requiring specialized technical knowledge or direct access to the \gls{llm} interface. This approach reduces technical barriers while maintaining high accuracy and reliability.

\begin{figure}
    \centering
    \includegraphics[width=0.8\linewidth]{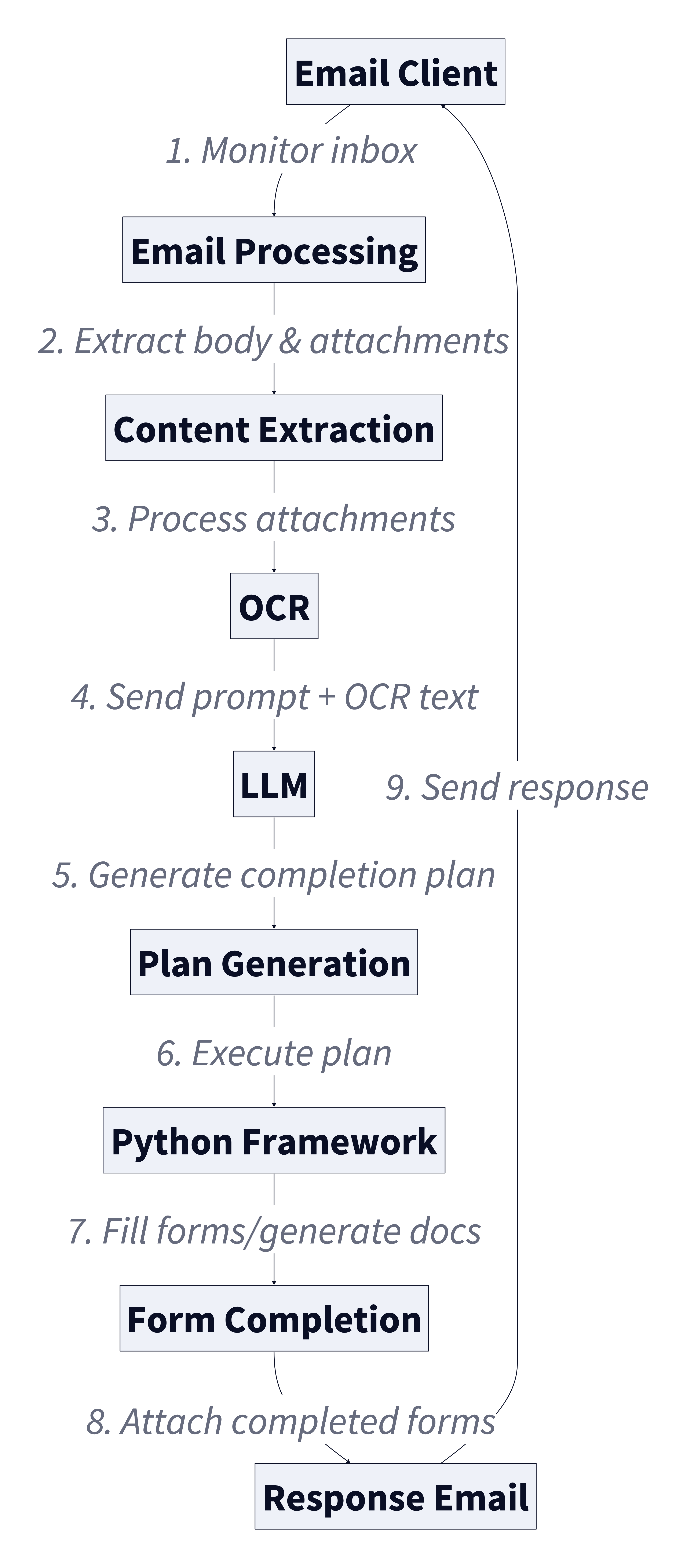}
    \caption{End-to-end workflow architecture for autonomous administrative form processing, demonstrating the coordinated operation of email parsing, document analysis, and automated form completion components.}
    \label{fig:proposed-workflow}
\end{figure}

Figure \ref{fig:proposed-workflow} illustrates the proposed workflow. This architecture serves as the foundation for testing various OCR and \gls{llm} algorithms while maintaining consistent system behavior and user experience.

The examples of the documents used were (1) \texttt{SOLICITUD SIMPLIFICADA DE: ALTA, BAJA O VARIACIÓN DE DATOS EN EL RÉGIMEN ESPECIAL DE AUTÓNOMO} \autocite{SeguridadSocial2025}, and (2) \texttt{SOLICITUD DE CONSULTA DE DOCUMENTOS DEL ARCHIVO GENERAL} \autocite{MinisterioJusticia2025}.

The full code of this proof of concept can be found at \href{https://github.com/andres-nav/smart-mail-buddy}{https://github.com/andres-nav/smart-mail-buddy}.

\section{Results}

Our evaluation focused on benchmarking the accuracy and reliability of various \glspl{llm} when integrated with AWS Rekognition for automated administrative form completion. The primary objective was to quantify model performance in extracting and correctly populating form fields, as well as to assess the processing efficiency of the combined \gls{ocr}-\gls{llm} pipeline.

\subsection{Performance Metrics}

The system was tested using seven different \glspl{llm}: \texttt{llama-4-maverick-17b-128e-instruct} \autocite{meta2025llama4maverick}, \texttt{llama-4-scout-17b-16e-instruct} \autocite{meta2025llama4scout}, \texttt{gemini-2.5-pro} \autocite{google_gemini2.5_2025}, \texttt{chatgpt-4.1} \autocite{openai_gpt4.1_2025}, \texttt{deepseek-r1} \autocite{deepseekai2025deepseekr1incentivizingreasoningcapability}, \texttt{llama-3.3-70b-instruct} \autocite{meta2024llama33}, and \texttt{qwen-qwq-32b} \autocite{qwq32b}. Each model was evaluated on a standardized form containing 29 fields. The evaluation criteria included the number of fields correctly filled, incorrectly filled, and left blank. The \gls{ocr} processing was performed using AWS Rekognition.

\begin{table}
\centering
\begin{tabular}{lcccc}
\hline
\textbf{Model} & \textbf{Correct} & \textbf{Incorrect} & \textbf{Blank} \\
\hline
Optimal result & 17 & 0 & 12 \\
llama-4-maverick-17b-128e-instruct & 16 & 2 & 11 \\
llama-4-scout-17b-16e-instruct & 14 & 3 & 12 \\
gemini-2.5-pro & 13 & 4 & 12 \\
chatgpt-4.1 & 12 & 5 & 12 \\
deepseek-r1 & 12 & 5 & 12 \\
llama-3.3-70b-instruct & 10 & 8 & 11 \\
qwen-qwq-32b & 10 & 10 & 9 \\
%TODO: test deepseek, claude, chatgpt & 1 & 1 & 1 \\
\hline
\end{tabular}
\caption{\gls{llm}-based form completion performance on a 29-field administrative form, comparing the number of fields each model filled correctly, incorrectly, or left blank. The "Optimal result" row represents the theoretical best outcome, where only fields for which information was provided in the input context are filled and all others are intentionally left blank, reflecting the ideal behavior of not hallucinating or fabricating data for fields lacking sufficient information. This benchmark highlights both the accuracy and the restraint of each model in form population, with blanks indicating appropriate omission rather than failure.}
\label{tab:llm_performance}
\end{table}

As shown in Table~\ref{tab:llm_performance}, the \texttt{llama-4} variants demonstrated superior performance, with \texttt{llama-4-maverick-17b-128e-instruct} achieving 16 correct fields (94\% of optimal benchmark). The analysis revealed three distinct performance tiers: top performers (16-14 correctly completed fields) consisted exclusively of \texttt{llama-4} family models; mid-range results (13-12 correct) included \texttt{gemini-2.5-pro}, \texttt{chatgpt-4.1}, and \texttt{deepseek-r1}; while lower accuracy outcomes (10 correct) featured smaller parameter models like \texttt{llama-3.3-70b-instruct} and \texttt{qwen-qwq-32b}. All evaluated systems exhibited non-trivial error rates (2-10 incorrect entries), underscoring the continued need for human verification in high-stakes administrative applications. 

Moreover, AWS Rekognition processed the entire document in 7.8 ±0.3 seconds, demonstrating the feasibility of real-time or near-real-time administrative automation workflows.

\subsection{Example Output}

\begin{figure}
\centering
    \begin{subfigure}{0.9\linewidth}
        \includegraphics[width=1\linewidth]{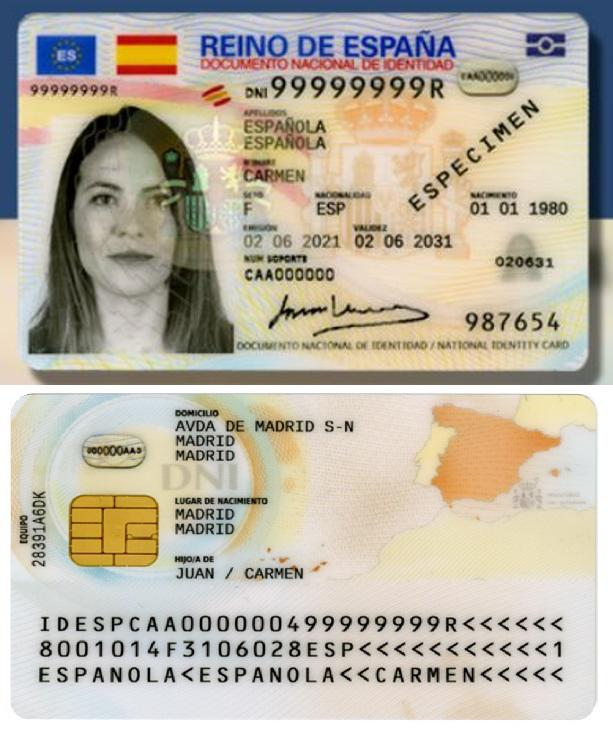}
        \caption{Example document of a Spanish National Identity Card of a fictitious person.}
        \label{fig:dni}
    \end{subfigure}

    \begin{subfigure}{0.9\linewidth}
        \includegraphics[width=1\linewidth]{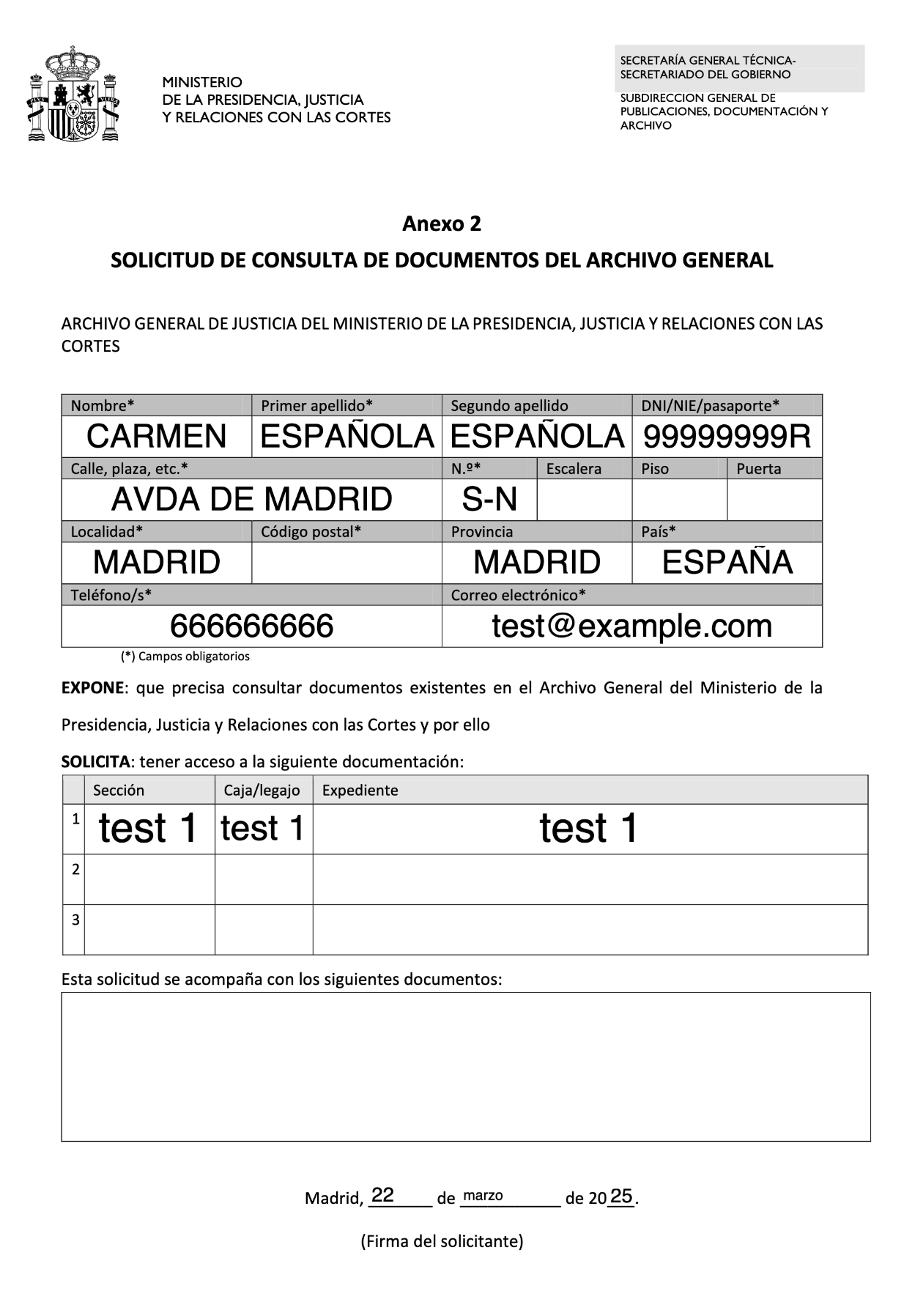}
        \caption{Filled form example of the \texttt{SOLICITUD DE CONSULTA DE DOCUMENTOS DEL ARCHIVO GENERAL} \autocite{MinisterioJusticia2025}.}
        \label{fig:form_filled}
    \end{subfigure}
    \caption{Example test performed with the proposed workflow where a form is filled with external documents using an AWS Rekognition to extract the text and a Llama 4 \gls{llm} to do the logic of filling the form.}
\end{figure}

An illustrative example of the completed form and the document used is provided in Figure~\ref{fig:form_filled}, highlighting the distribution of correct, incorrect, and blank fields for the top-performing \texttt{llama-4-maverick-17b-128e-instruct}, where more than half of the fields were correctly filled, with minimal errors. This demonstrates the practical viability of the system for semi-automated administrative processing, though further optimization is required to reduce the number of blank and incorrect fields.

\subsection{Comparison to Manual Processing}

There is limited published research quantifying the exact time required for administrative staff to manually complete complex government forms, as the duration can vary significantly depending on the form's complexity and the staff's familiarity with the process. In our practical tests, completing the selected form manually required between 15 and 20 minutes per instance. By contrast, our automated system completed the same form in under 10 seconds, 7.8 seconds. However, we include a human supervision and validation time of approximately 5 minutes per document to validate its correctness. This approach reduces the active labor required by administrative staff to a brief review, rather than full manual entry.

To contextualize the economic impact, we estimate costs using the median annual gross salary plus the social security taxes in Spain, which is approximately €40,000~\autocite{ine2023wages}. Assuming a 40-hour work week over 52 weeks, this translates to an hourly wage of roughly €19.23. The cost for one manual form completion, assuming 15 minutes per form,  is therefore about €4.81. With our system, the cost per form consists of the maximum estimated cloud processing cost (€0.10 per form, based on our simulations) and the human supervision cost (5 minutes, or €1.60). The total cost per form using the system is thus €1.70, yielding a savings of €3.11 per form, or a 64.6\% reduction in direct processing costs.

\begin{table}
    \centering
    \begin{tabular}{lccc}
        \hline
        \textbf{Method} & \textbf{Time [min]} & \textbf{Cost [€]} \\
        \hline
        Manual Completion & 15 & 4.81 \\
        System + Supervision & 5 & 1.70 \\
        \hline
    \end{tabular}
    \caption{Comparison of estimated time and cost required to complete a single administrative form manually versus using the automated system with human supervision.}
    \label{tab:cost_comparison}
\end{table}

Table~\ref{tab:cost_comparison} summarizes the estimated time and cost per form for both manual and automated processing methods. In terms of operational efficiency, our system reduces the time required for form processing from the typical 15-20 minutes needed for manual completion to just 7.8 seconds for automated extraction and field population. Even when including an additional 5 minutes for human supervision and validation, the total staff time per form is reduced by a factor of three to four compared to manual processing. This substantial reduction in time directly translates into lower labor costs. It enables administrative staff to focus on more complex or value-added activities. The cost savings are particularly significant for organizations handling large volumes of administrative forms, where automation not only improves efficiency but also ensures greater consistency and reliability in document handling. 

\section{Conclusions}

This research demonstrates the significant potential of combining \glspl{llm} with email interfaces to create accessible automation solutions for administrative tasks. By leveraging email as a familiar interface, we remove technical barriers that might otherwise limit \gls{llm} adoption among administrative staff. The proposed architecture effectively integrates email clients, \gls{ocr} technology, \glspl{llm}, and Python automation to create a comprehensive solution for automating administrative tasks.

Our proof-of-concept implementation and evaluation show that this approach enables non-technical users to harness advanced \gls{llm} capabilities without requiring new technical skills or workflow changes. The system completed complex administrative forms by extracting and populating information from attached documents, with the top-performing model (\texttt{llama-4-maverick-17b-128e-instruct}) correctly filling 16 out of 29 fields. Automated processing using AWS Rekognition and \gls{llm} inference required just 7.8 seconds per document, with an additional 5 minutes for human supervision. This workflow reduced the total staff time required by a factor of three to four compared to manual completion.

Cost analysis further highlights the benefits of this approach. The estimated cost per form using the automated system, including cloud processing and human supervision, was €1.70, a 64\% reduction compared to the €4.81 per form for manual processing. These results demonstrate that the proposed system not only increases efficiency and reduces repetitive workload for administrative staff but also offers substantial cost savings, particularly for organizations processing large volumes of forms.

Despite these promising results, our system faces challenges similar to those identified in healthcare \gls{llm} implementations, including occasional inaccuracies, potential privacy concerns, and the need for human oversight in sensitive contexts. These limitations highlight the importance of designing systems that augment rather than replace human administrative staff, ensuring that automation enhances human capabilities while maintaining appropriate oversight and quality control mechanisms.

The integration of \glspl{llm} with email interfaces represents a practical approach to automating administrative tasks. By building on familiar communication channels and focusing on high-volume, repetitive tasks, our system demonstrates how organizations can begin capturing value from \gls{llm} technologies without requiring extensive retraining or workflow disruption. This work establishes a foundation for more widespread adoption of \gls{ai}-powered administrative automation while maintaining the human-centered design principles essential for successful technology integration.

\section{Future Work}

One important direction for future work is to add support for \glspl{mcp}. They are a new standard that makes it easier to connect large language models with other tools, data sources, and enterprise systems. Using \glspl{mcp} could help the system work better with different software and automate more complex workflows. However, when this paper was written, \glspl{mcp} were not yet available or mature enough to use in practice, so they were not included in our design.

Other areas to improve include supporting more document formats, such as Word files, presentations, and spreadsheets, which would make the system more useful in different administrative settings. This will require building new modules to handle each type of file. Finally, connecting the system with enterprise platforms like \gls{crm}, \gls{erp}, or \gls{hris} remains a key goal, as this would allow for full process automation across an organization while keeping the easy-to-use email interface for users.
.
\section*{Acknowledgment}

The authors would like to acknowledge the support of Spanish projects 6G-INTEGRATION-3 (TSI-063000-2021-127) and Fun4Date-Redes (PID2022-136684OB-C21).


@inproceedings{10.1145/3636534.3649379,
author = {Wen, Hao and Li, Yuanchun and Liu, Guohong and Zhao, Shanhui and Yu, Tao and Li, Toby Jia-Jun and Jiang, Shiqi and Liu, Yunhao and Zhang, Yaqin and Liu, Yunxin},
title = {AutoDroid: LLM-powered Task Automation in Android},
year = {2024},
isbn = {9798400704895},
publisher = {Association for Computing Machinery},
address = {New York, NY, USA},
url = {https://doi.org/10.1145/3636534.3649379},
doi = {10.1145/3636534.3649379},
booktitle = {Proceedings of the 30th Annual International Conference on Mobile Computing and Networking},
pages = {543–557},
numpages = {15},
location = {Washington D.C., DC, USA},
series = {ACM MobiCom '24}
}

@INPROCEEDINGS{10527275,
  author={Gebreab, Senay A. and Salah, Khaled and Jayaraman, Raja and Habib ur Rehman, Muhammad and Ellaham, Samer},
  booktitle={2024 12th International Symposium on Digital Forensics and Security (ISDFS)}, 
  title={LLM-Based Framework for Administrative Task Automation in Healthcare}, 
  year={2024},
  volume={},
  number={},
  pages={1-7},
  doi={10.1109/ISDFS60797.2024.10527275}}

@article{Peddinti2023,
  author    = {Sudhakar Reddy Peddinti and Subba Rao Katragadda and Brij Kishore Pandey and Ajay Tanikonda},
  title     = {Utilizing Large Language Models for Advanced Service Management: Potential Applications and Operational Challenges},
  journal   = {Journal of Science \& Technology},
  volume    = {4},
  number    = {2},
  pages     = {},
  year      = {2023},
  month     = {March--April},
  note      = {Available at SSRN: \url{https://ssrn.com/abstract=5119925} or \url{http://dx.doi.org/10.2139/ssrn.5119925}},
  url       = {https://ssrn.com/abstract=5119925},
  doi       = {10.2139/ssrn.5119925}
}

@article{Pesch_2025, title={Potentials and Challenges of Large Language Models (LLMs) in the Context of Administrative Decision-Making}, volume={16}, DOI={10.1017/err.2024.99}, number={1}, journal={European Journal of Risk Regulation}, author={Pesch, Paulina Jo}, year={2025}, pages={76–95}}

@article{21068f9715904033900a91cad6b9f758,
title = "The digital workplace is key to digital innovation",
author = "Kristine Dery and Sebastian, {Ina M.} and {van der Meulen}, Nick",
year = "2017",
language = "English",
volume = "16",
pages = "135--152",
journal = "MIS Quarterly Executive",
issn = "1540-1960",
publisher = "Indiana University Press",
number = "2",
}

@article{ess2025ai,
  author    = {Ess, SA and Mackey, AJ and Yarowsky, DE},
  title     = {Artificial Intelligence Scribe and Large Language Model Technology in Healthcare Documentation: Advantages, Limitations, and Recommendations},
  journal   = {Plastic and Reconstructive Surgery Global Open},
  year      = {2025},
  volume    = {13},
  number    = {1},
  pages     = {e6450},
  month     = {Jan 16},
  doi       = {10.1097/GOX.0000000000006450},
  pmid      = {39823022},
  pmcid     = {PMC11737491}
}

@misc{chen2025empiricalstudychallengesllm,
      title={An Empirical Study on Challenges for LLM Application Developers}, 
      author={Xiang Chen and Chaoyang Gao and Chunyang Chen and Guangbei Zhang and Yong Liu},
      year={2025},
      eprint={2408.05002},
      archivePrefix={arXiv},
      primaryClass={cs.SE},
      url={https://arxiv.org/abs/2408.05002}, 
}

@misc{kong2023organizationalbulkemailsystems,
      title={Organizational Bulk Email Systems: Their Role and Performance in Remote Work}, 
      author={Ruoyan Kong and Haiyi Zhu and Joseph A. Konstan},
      year={2023},
      eprint={2308.05085},
      archivePrefix={arXiv},
      primaryClass={cs.HC},
      url={https://arxiv.org/abs/2308.05085}, 
}

@misc{josephs2024communicationnetworkdynamicslarge,
      title={Communication network dynamics in a large organizational hierarchy}, 
      author={Nathaniel Josephs and Sida Peng and Forrest W. Crawford},
      year={2024},
      eprint={2208.01208},
      archivePrefix={arXiv},
      primaryClass={stat.AP},
      url={https://arxiv.org/abs/2208.01208}, 
}

@inproceedings{Khare_2022,
   title={E-Mail Assistant – Automation of E-Mail Handling and Management using Robotic Process Automation},
   url={http://dx.doi.org/10.1109/DASA54658.2022.9765017},
   DOI={10.1109/dasa54658.2022.9765017},
   booktitle={2022 International Conference on Decision Aid Sciences and Applications (DASA)},
   publisher={IEEE},
   author={Khare, Arpit and Singh, Sudhakar and Mishra, Richa and Prakash, Shiv and Dixit, Pratibha},
   year={2022},
   month=mar, pages={511–516} }

@misc{thiergart2021understandingemailsdraftingresponses,
      title={Understanding Emails and Drafting Responses -- An Approach Using GPT-3}, 
      author={Jonas Thiergart and Stefan Huber and Thomas Übellacker},
      year={2021},
      eprint={2102.03062},
      archivePrefix={arXiv},
      primaryClass={cs.AI},
      url={https://arxiv.org/abs/2102.03062}, 
}

@misc{kannan2016smartreplyautomatedresponse,
      title={Smart Reply: Automated Response Suggestion for Email}, 
      author={Anjuli Kannan and Karol Kurach and Sujith Ravi and Tobias Kaufmann and Andrew Tomkins and Balint Miklos and Greg Corrado and Laszlo Lukacs and Marina Ganea and Peter Young and Vivek Ramavajjala},
      year={2016},
      eprint={1606.04870},
      archivePrefix={arXiv},
      primaryClass={cs.CL},
      url={https://arxiv.org/abs/1606.04870}, 
}

@misc{bose2019fieldlabelpredictionautofill,
      title={Field Label Prediction for Autofill in Web Browsers}, 
      author={Joy Bose},
      year={2019},
      eprint={1912.08809},
      archivePrefix={arXiv},
      primaryClass={cs.HC},
      url={https://arxiv.org/abs/1912.08809}, 
}

@misc{aveni2023omnifilldomainagnosticformfilling,
      title={OmniFill: Domain-Agnostic Form Filling Suggestions Using Multi-Faceted Context}, 
      author={Timothy J. Aveni and Armando Fox and Björn Hartmann},
      year={2023},
      eprint={2310.17826},
      archivePrefix={arXiv},
      primaryClass={cs.HC},
      url={https://arxiv.org/abs/2310.17826}, 
}

@misc{amazonworkmail,
  title        = {Amazon WorkMail},
  howpublished = {\url{https://aws.amazon.com/workmail/}},
  note         = {Accessed: 2025-05-18},
  year         = {2025},
  author       = {{Amazon Web Services, Inc.}}
}

@misc{tesseract2025,
  title = {Tesseract Open Source OCR Engine},
  author = {Smith, Ray and Podobny, Zdenko and {Tesseract OCR Contributors}},
  year = {2025},
  url = {https://github.com/tesseract-ocr/tesseract},
  note = {Version 5.5.1},
  organization = {Google Inc.},
  howpublished = {\url{https://github.com/tesseract-ocr/tesseract}}
}

@misc{easyocr,
  author       = {JaidedAI},
  title        = {EasyOCR: Ready-to-use OCR with 80+ Languages Supported},
  year         = {2020},
  howpublished = {\url{https://github.com/JaidedAI/EasyOCR}},
  note         = {Accessed: 2025-05-29}
}

@misc{amazon_textract,
  author       = {{Amazon Web Services}},
  title        = {Amazon Textract: Automatically Extract Printed Text, Handwriting, Layout Elements, and Data from Any Document},
  year         = {2018},
  howpublished = {\url{https://aws.amazon.com/textract/}},
  note         = {Accessed: 2025-05-29}
}

@misc{amazon_rekognition,
  author       = {{Amazon Web Services}},
  title        = {Amazon Rekognition: Automate and Lower the Cost of Your Image Recognition and Video Analysis with ML},
  year         = {2016},
  howpublished = {\url{https://aws.amazon.com/rekognition/}},
  note         = {Accessed: 2025-05-29}
}

@misc{kulkarni2025agentsllmagenticworkflow,
      title={Agent-S: LLM Agentic workflow to automate Standard Operating Procedures}, 
      author={Mandar Kulkarni},
      year={2025},
      eprint={2503.15520},
      archivePrefix={arXiv},
      primaryClass={cs.HC},
      url={https://arxiv.org/abs/2503.15520}, 
}

@misc{SeguridadSocial2025,
    title = {TA.0521.1 Simplificada (V9)},
    author = {Ministerio de Justicia},
    url = {https://sede.seg-social.gob.es/wps/wcm/connect/sede/9b1bd995-f5f7-4dba-aba1-3a8e04105d79/TA_0521_1%2BSimplificada%2B%28V9%29.pdf?MOD=AJPERES},
    note = {Formulario oficial de la Seguridad Social de España},
    institution = {Seguridad Social, Gobierno de España},
    year = {2025},
    urldate = {2025-06-08}
}

@misc{MinisterioJusticia2025,
    title = {Formulario de solicitud de consulta de documentos del Archivo General},
    author = {Ministerio de Justicia},
    url = {https://sede.mjusticia.gob.es/es/TramitesSede/Documents/1292429619664-Formulario_de_solicitud_de_consulta.pdf},
    note = {Formulario oficial del Ministerio de Justicia de España},
    institution = {Ministerio de Justicia, Gobierno de España},
    year = {2025},
    urldate = {2025-06-08}
}

@online{meta2025llama4maverick,
  title        = {Llama 4 Maverick 17B-128E Instruct},
  author       = {Meta AI},
  year         = {2025},
  url = {https://huggingface.co/meta-llama/Llama-4-Maverick-17B-128E-Instruct},
  note         = {Model release date: April 5, 2025. Llama 4 Maverick is a 17B parameter, 128-expert, natively multimodal large language model released under the Llama 4 Community License. Knowledge cutoff: August 2024.}
}

@online{meta2025llama4scout,
  title        = {Llama 4 Scout 17B-16E Instruct},
  author       = {Meta AI},
  year         = {2025},
  url = {https://huggingface.co/meta-llama/Llama-4-Scout-17B-16E-Instruct},
  note         = {Model release date: April 5, 2025. Llama 4 Scout is a 17B parameter, 16-expert, natively multimodal large language model released under the Llama 4 Community License. Knowledge cutoff: August 2024.}
}

@online{meta2024llama33,
  title        = {Llama 3.3 70B Instruct},
  author       = {Meta AI},
  year         = {2024},
  url = {https://huggingface.co/meta-llama/Llama-3.3-70B-Instruct},
  note         = {Model release date: December 6, 2024. Llama 3.3 is a 70B parameter, instruction-tuned, multilingual large language model released under the Llama 3.3 Community License. Knowledge cutoff: December 2023.}
}

@online{qwq32b,
  title  = {QwQ-32B: Embracing the Power of Reinforcement Learning},
  author = {Qwen Team},
  year   = {2025},
  month  = {March},
  url    = {https://huggingface.co/Qwen/QwQ-32B}
}

@online{ine2023wages,
  author = {{Instituto Nacional de Estadística (INE)}},
  title = {Wage Structure Survey (Encuesta de Estructura Salarial)},
  year = {2023},
  url = {https://www.ine.es/dyngs/INEbase/en/operacion.htm?c=Estadistica_C&cid=1254736177025&menu=ultiDatos&idp=1254735976596},
  note = {Average annual salary in 2023: 28,049.94 euros per worker. Published: 28/05/2025.},
  organization = {Instituto Nacional de Estadística (INE)}
}

@online{google_gemini2.5_2025,
title = {Gemini 2.5: Our most intelligent AI model},
author = {{Google DeepMind}},
year = {2025},
month = {March},
url = {https://blog.google/technology/google-deepmind/gemini-model-thinking-updates-march-2025/},
note = {Last updated March 26. Introduces Gemini 2.5, a state-of-the-art "thinking model" designed for advanced reasoning and complex problem-solving.}
}

@online{openai_gpt4.1_2025,
title = {Introducing GPT-4.1 in the API},
author = {{OpenAI}},
year = {2025},
month = {June},
url = {https://openai.com/index/gpt-4-1/},
note = {Announces the release of GPT-4.1, GPT-4.1 mini, and GPT-4.1 nano, with major improvements in coding, instruction following, and long context handling.}
}

@misc{deepseekai2025deepseekr1incentivizingreasoningcapability,
      title={DeepSeek-R1: Incentivizing Reasoning Capability in LLMs via Reinforcement Learning}, 
      author={DeepSeek-AI and Daya Guo and Dejian Yang and Haowei Zhang and Junxiao Song and Ruoyu Zhang and Runxin Xu and Qihao Zhu and Shirong Ma and Peiyi Wang and Xiao Bi and Xiaokang Zhang and Xingkai Yu and Yu Wu and Z. F. Wu and Zhibin Gou and Zhihong Shao and Zhuoshu Li and Ziyi Gao and Aixin Liu and Bing Xue and Bingxuan Wang and Bochao Wu and Bei Feng and Chengda Lu and Chenggang Zhao and Chengqi Deng and Chenyu Zhang and Chong Ruan and Damai Dai and Deli Chen and Dongjie Ji and Erhang Li and Fangyun Lin and Fucong Dai and Fuli Luo and Guangbo Hao and Guanting Chen and Guowei Li and H. Zhang and Han Bao and Hanwei Xu and Haocheng Wang and Honghui Ding and Huajian Xin and Huazuo Gao and Hui Qu and Hui Li and Jianzhong Guo and Jiashi Li and Jiawei Wang and Jingchang Chen and Jingyang Yuan and Junjie Qiu and Junlong Li and J. L. Cai and Jiaqi Ni and Jian Liang and Jin Chen and Kai Dong and Kai Hu and Kaige Gao and Kang Guan and Kexin Huang and Kuai Yu and Lean Wang and Lecong Zhang and Liang Zhao and Litong Wang and Liyue Zhang and Lei Xu and Leyi Xia and Mingchuan Zhang and Minghua Zhang and Minghui Tang and Meng Li and Miaojun Wang and Mingming Li and Ning Tian and Panpan Huang and Peng Zhang and Qiancheng Wang and Qinyu Chen and Qiushi Du and Ruiqi Ge and Ruisong Zhang and Ruizhe Pan and Runji Wang and R. J. Chen and R. L. Jin and Ruyi Chen and Shanghao Lu and Shangyan Zhou and Shanhuang Chen and Shengfeng Ye and Shiyu Wang and Shuiping Yu and Shunfeng Zhou and Shuting Pan and S. S. Li and Shuang Zhou and Shaoqing Wu and Shengfeng Ye and Tao Yun and Tian Pei and Tianyu Sun and T. Wang and Wangding Zeng and Wanjia Zhao and Wen Liu and Wenfeng Liang and Wenjun Gao and Wenqin Yu and Wentao Zhang and W. L. Xiao and Wei An and Xiaodong Liu and Xiaohan Wang and Xiaokang Chen and Xiaotao Nie and Xin Cheng and Xin Liu and Xin Xie and Xingchao Liu and Xinyu Yang and Xinyuan Li and Xuecheng Su and Xuheng Lin and X. Q. Li and Xiangyue Jin and Xiaojin Shen and Xiaosha Chen and Xiaowen Sun and Xiaoxiang Wang and Xinnan Song and Xinyi Zhou and Xianzu Wang and Xinxia Shan and Y. K. Li and Y. Q. Wang and Y. X. Wei and Yang Zhang and Yanhong Xu and Yao Li and Yao Zhao and Yaofeng Sun and Yaohui Wang and Yi Yu and Yichao Zhang and Yifan Shi and Yiliang Xiong and Ying He and Yishi Piao and Yisong Wang and Yixuan Tan and Yiyang Ma and Yiyuan Liu and Yongqiang Guo and Yuan Ou and Yuduan Wang and Yue Gong and Yuheng Zou and Yujia He and Yunfan Xiong and Yuxiang Luo and Yuxiang You and Yuxuan Liu and Yuyang Zhou and Y. X. Zhu and Yanhong Xu and Yanping Huang and Yaohui Li and Yi Zheng and Yuchen Zhu and Yunxian Ma and Ying Tang and Yukun Zha and Yuting Yan and Z. Z. Ren and Zehui Ren and Zhangli Sha and Zhe Fu and Zhean Xu and Zhenda Xie and Zhengyan Zhang and Zhewen Hao and Zhicheng Ma and Zhigang Yan and Zhiyu Wu and Zihui Gu and Zijia Zhu and Zijun Liu and Zilin Li and Ziwei Xie and Ziyang Song and Zizheng Pan and Zhen Huang and Zhipeng Xu and Zhongyu Zhang and Zhen Zhang},
      year={2025},
      eprint={2501.12948},
      archivePrefix={arXiv},
      primaryClass={cs.CL},
      url={https://arxiv.org/abs/2501.12948}, 
}
\end{document}